\documentclass[onecolumn,epjc3]{svjour3}
\smartqed  
\RequirePackage{graphicx}
\journalname{Eur. Phys. J. C}
\begin{document}

\title{A new resummation scheme in scalar field theories
} 


\author{Wei-jie Fu\thanksref{e1,addr1,addr2}
}

\thankstext{e1}{e-mail: fuw@brandonu.ca}


\institute{Department of Physics, Brandon University, Brandon,
Manitoba, R7A 6A9 Canada \label{addr1}
           \and
           Winnipeg Institute for Theoretical Physics, Winnipeg, Manitoba
\label{addr2}
}

\date{Received: date / Accepted: date}

\maketitle

\begin{abstract}
A new resummation scheme in scalar field theories is proposed by
combining parquet resummation techniques and flow equations, which
is characterized by a hierarchy structure of the Bethe--Salpeter
(BS) equations. The new resummation scheme greatly improves on the
approximations for the BS kernel. Resummation of the BS kernel in
the $t$ and $u$ channels to infinite order is equivalent to truncate
the effective action to infinite order. Our approximation approaches
ensure that the theory can be renormalized, which is very important
for numerical calculations. Two-point function can also be obtained
from the four-point one through flow evolution equations resulting
from the functional renormalization group. BS equations of different
hierarchies and the flow evolution equation for the propagator
constitute a closed self-consistent system, which can be solved
completely. 
\end{abstract}

\section{Introduction}
\label{intro}

The Bethe-Salpeter (BS) equation plays an important role in
many-body theories. It has been widely used in various research
fields, e.g. the strongly correlated electron
systems~\cite{Bickers1989a,Bickers1989b}, hadron
physics~\cite{Maris1997,Bender1996,Maris2003,Chang2009}, particle
physics, and the field
theory~\cite{van2002,Blaizot2004,Berges2005,Aarts2004}. The BS
equation resums its kernel, usually called the BS kernel, to
infinite order. Therefore, non-perturbative effects are included in
the BS equation. However, usually in actual applications of the BS
equation, it is impossible to obtain an exact kernel and we have to
approximate it. For example, when the BS equation is employed to
study the properties of mesons in the QCD, the usually adopted
approximation is called the rainbow-ladder approximation, in which
bare quark--gluon vertices are used to construct the four-quark
kernel~\cite{Bender1996}. However, the rainbow-ladder approximation
is the simplest approximation, which loses lots of information. Lots
of efforts have been made to go beyond the rainbow-ladder
approximation~\cite{Bender1996,Chang2009}.

In fact, when we employ the BS equation to describe a many-body
system, the central point is whether the kernel used is good enough
to obtain what are observed in experiments. If it is not enough, how
can we improve it? In this work, we will discuss how to improve the
approximations for the BS kernel systematically. It is found that a
new resummation scheme can be used to greatly improve the
approximations by employing the hierarchy structure of the BS
equation.

In this work we will work under the formalism of the two-particle
irreducible (2PI) effective action theory~\cite{Cornwall1974}, which
is also known as the $\Phi$-derivable approximations in the
condensed matter physics~\cite{Luttinger1960,Baym1962}. In the
recent years, the 2PI effective action theory has attracted a lot of
attention. It has been applied to calculate the entropy of the
quark--gluon plasma and other thermodynamic
quantities~\cite{Blaizot1999,Berges2005a}, describe the
non-equilibrium dynamics with subsequent late-time
thermalization\cite{Berges2001}, compute the shear viscosity in the
thermal field theory~\cite{Aarts2004}, and so on. In particular, we
would like to emphasize that due to many people's
contributions~\cite{van2002,Blaizot2004,Berges2005,Reinosa2010}, it
has been clear that the 2PI effective action theory can be
renormalized, which is quite non-trivial for a non-perturbative
approach. Therefore, in our following discussions, the
renormalizability of our approximation approaches will work as a
prerequisite demand. The renormalizability should not be violated at
any case.

The paper is organized as follows. In Sect. \ref{BSsection} we
discuss the hierarchy structure of the BS equation and how to
improve on the approximations for the BS kernel by employing this
hierarchy structure. In Sect. \ref{propagatorsection} we show how
the propagator and the effective action are obtained from the
four-point vertex through flow evolution equations resulting from
the functional renormalization
group~\cite{Wetterich1993,Blaizot2011}. Section \ref{summary}
summarizes the results and gives outlooks.

\section{Hierarchy Structure of the BS Equation}
\label{BSsection} \vspace{5pt}

We consider the following scalar field theory with a non-local
regulator term
\begin{equation}
S_{\kappa}[\varphi]=S[\varphi]+\Delta
S_{\kappa}[\varphi],\label{Eq1}
\end{equation}
with
\begin{eqnarray}
S[\varphi]&=&\frac{1}{2}\varphi_{i}iG_{0,ij}^{-1}\varphi_{j}-\frac{\lambda
}{4!}\varphi^{4},\\
\Delta
S_{\kappa}[\varphi]&=&-\frac{1}{2}\varphi_{i}R_{\kappa,ij}\varphi_{j},\label{Eq3}
\end{eqnarray}
where $iG_{0,ij}^{-1}=(-\partial^{2}-m^{2})\delta_{ij}$, Here
summations or integrals are assumed for repeated indices. The
non-local term $\Delta S_{\kappa}$ in Eq.~(\ref{Eq3}) is employed to
suppress quantum fluctuations whose wave lengths are larger than
$1/\kappa$, where $\kappa$ has a dimension of
momentum~\cite{Wetterich1993}. This is achieved as follows: the
non-local term in momentum space is given by
\begin{equation}
\Delta
S_{\kappa}[\varphi]=-\frac{1}{2}\int\frac{d^{D}q}{(2\pi)^{D}}R_{\kappa}(q)
\varphi(q)\varphi(-q).
\end{equation}
The regulator $R_{\kappa}(q)$ is chosen to have following
properties: when $q\ll \kappa$, $R_{\kappa}(q)\sim \kappa^{2}$, then
the non-local term becomes a mass term with large mass $\kappa$,
which suppresses quantum fluctuations with wave lengths $1/q\gg
1/\kappa$; when $q\geq\kappa$, $R_{\kappa}(q)\rightarrow 0$, so
fluctuations with wave lengths $1/q\leq 1/\kappa$ are not affected.
In the functional renormalization group theory, beginning from a
classical action at an ultraviolet scale $\Lambda$, one can obtain
the corresponding quantum action which takes into account all
quantum fluctuations through the evolution of flow equations from
$\kappa=\Lambda$ to 0.

From Eq.~(\ref{Eq1}) we can obtain the corresponding 2PI effective
action. The generating functional with one- and two-point sources is
given by
\begin{eqnarray}
Z_{\kappa}[J,J_{2}]&=&\int[d\varphi]\exp\bigg\{i\Big(S_{\kappa}[\varphi]+J_{i}\varphi_{i}
+\frac{1}{2}\varphi_{i}J_{2,ij}\varphi_{j}\Big)\bigg\},\\
W_{\kappa}[J,J_{2}]&=&-i\ln Z_{\kappa}[J,J_{2}].
\end{eqnarray}
Performing the functional derivative with respect to sources, one
obtains
\begin{eqnarray}
\frac{\delta W_{\kappa}}{\delta J_{i}}
&=&\langle \varphi_{i}\rangle\equiv\phi_{i},\\
\frac{\delta W_{\kappa}}{\delta
J_{2,ij}}&=&\frac{1}{2}(\phi_{i}\phi_{j}+G_{ij}),
\end{eqnarray}
where $\phi$ is the expected value of the field $\varphi$ and $G$ is
the full propagator. The 2PI effective action is obtained through
the Legendre transformation from $W_{\kappa}[J,J_{2}]$, i.e.,
\begin{eqnarray}
\Gamma_{\kappa}[\phi,G]&=&W_{\kappa}-J_{i}\frac{\delta
W_{\kappa}}{\delta J_{i}} -J_{2,ij}\frac{\delta W_{\kappa}}{\delta
J_{2,ij}}\nonumber \\
&=&W_{\kappa}-J_{i}\phi_{i}-\frac{1}{2}J_{2,ij}(\phi_{i}\phi_{j}+G_{ij}).
\end{eqnarray}
Expressed in terms of $\phi$ and $G$, the 2PI effective action
reads~\cite{Cornwall1974}
\begin{eqnarray}
\Gamma_{\kappa}[\phi,G]&=&\frac{1}{2}\phi_{i}iG_{0\kappa,ij}^{-1}\phi_{j}+\frac{i}{2}\mathrm{Tr}\ln
G^{-1}
+\frac{i}{2}\mathrm{Tr}G_{0\kappa}^{-1}G+\Gamma_{\mathrm{int}}[\phi,G],\label{Gamma_2PI}
\end{eqnarray}
where we have $iG_{0\kappa}^{-1}=iG_{0}^{-1}-R_{\kappa}$ and the
interacting part of the effective action is
\begin{eqnarray}
\Gamma_{\mathrm{int}}[\phi,G]&=&\frac{1}{2}\phi_{i}i\delta
G_{0,ij}^{-1}\phi_{j} +\frac{i}{2}\mathrm{Tr}\delta
G_{0}^{-1}G-\frac{\lambda+\delta\lambda}{4!}\phi^{4}\nonumber
\\&&-\frac{\lambda+\delta\lambda}{4}\phi^{2}G+\Gamma_{2}[\phi,G],\label{Gamma_int}
\end{eqnarray}
where $\Gamma_{2}$ is given by all 2PI vacuum graphs whose vertices
are given by the terms cubic or quartic in $\varphi$ in the
expanding expression of $S[\phi+\varphi]-S[\phi]$, and propagators
are the full ones. We have employed renormalized quantities in
Eq.~(\ref{Gamma_2PI}) and Eq.~(\ref{Gamma_int}). They are related to
the bare quantities (with subscript B) through the following
relations:
\begin{eqnarray}
\label{Zdefn} && \delta m^{2} = Zm_{B}^{2}-m^{2}\,,~~~\delta \lambda
= Z^{2}\lambda_{B}-\lambda\,,~~~\delta Z =Z-1\,.\\
&& ZG^{-1}_{0B} =  G^{-1}_{0}+ \delta G^{-1}_{0}\,,~~~\delta
G^{-1}_{0}=i(\delta Z \partial^{2}+\delta m^2)\,,~~~ G_{B} =
ZG\,.\nonumber
\end{eqnarray}

In actual calculations, we must make approximations. In another
words, we have to truncate the 2PI vacuum graphs included in the
$\Gamma_{2}$ in Eq.~(\ref{Gamma_int}). For example, one can expand
$\Gamma_{2}$ in order of loops of skeletons, which is also known as
the $\Phi$ derivable approximation; one can also expand $\Gamma_{2}$
in order of $1/N$ in the $O(N)$ model. Whatever it is, the common
feature of these approximation approaches is that approximations are
made to the effective action. Once the approximations are made, we
can employ the approximative effective action to obtain the
self-consistent equation for the two-point function. We can also
obtain the BS equation for the four-point function.

In this work, we will not go the same way as those described above.
We will reverse the procedure, i.e., first, we make an approximation
to the BS equation, then return to the two-point function and the
effective potential. Comparing our approach with the conventional
one, we find that our approach has several advantages: First, our
approach makes much more powerful approximations. Second, we can
easily observe the hierarchy structure of the BS equations from our
approach. Finally, one can employ the hierarchy structure of the BS
equations to reorganize the infinite diagrams through resummation.
Same as the conventional approach, our approach produces
approximations which are always renormalizable. As for a
non-perturbative approximation, renormalizability is a quite
non-trivial demand.

\begin{figure}[!htb]
\includegraphics[scale=0.75]{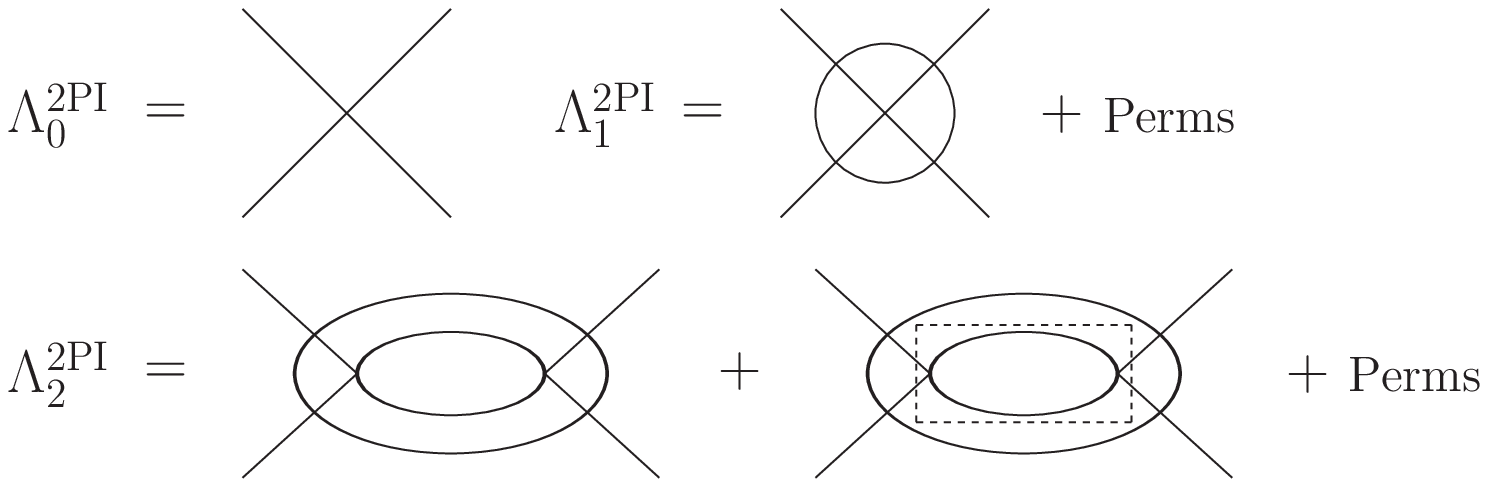}
\caption{Lowest three order contributions to the diagrams which are
2PI in all the three channels.}\label{f4}
\end{figure}

In order to simplify our calculations but without lose of the
generality, we consider the symmetric case in the following
discussions, i.e., $\phi=0$ and 3-point vertex is also vanishing. We
begin with the BS equation for the four-point vertex in the
coordinate space:
\begin{equation}
M_{ij;kl}=\Lambda_{ij;kl}+\frac{1}{2}\Lambda_{ij;k'l'}G_{k'i'}G_{l'j'}M_{i'j';kl},\label{BS1}
\end{equation}
where the kernel is given by
\begin{equation}
\Lambda_{ij;kl}=4i\frac{\delta^{2}\Gamma_{\mathrm{int}}}{\delta
G_{ij}\delta G_{kl}}.\label{}
\end{equation}
We know that the BS equation resums diagrams in one channel to
infinite order. Here we designate this channel as $s$ channel. It is
also known that the renormalizability of the BS equation demands
that in the kernel $\Lambda$, the two-particle reducible diagrams
only can appear in the $t$ or $u$ channels. They are prohibited in
the $s$ channel~\cite{van2002,Blaizot2004,Berges2005}. As we have
said, the $s$ channel is resummed to infinite order through the BS
equation. Therefore, it is natural to remind us of the question: can
we resum the two other channels to infinite order as well? In fact,
inspired by the parquet resummation techniques~\cite{Roulet1969}, we
realize that this can be obtained by employing the BS equation once
more, i.e., the kernel of the BS equation in Eq.~(\ref{BS1}) can be
constructed with the solution of another BS equation. Therefore, we
can express the kernel as
\begin{equation}
\Lambda_{ij;kl}=(\Lambda_{ij;kl}^{\mathrm{2PI}}-i\delta\lambda\delta_{ij}\delta_{ik}\delta_{il})
+(M'_{ik;jl}-\Lambda'_{ik;jl})+(M'_{il;jk}-\Lambda'_{il;jk}),\label{Lamb}
\end{equation}
and we have
\begin{eqnarray}
M'_{ik;jl}&=&\Lambda'_{ik;jl}+\frac{1}{2}\Lambda'_{ik;i_{1}i_{2}}G_{i_{1}j_{1}}G_{i_{2}j_{2}}M'_{j_{1}j_{2};jl}, \label{BS2}\\
M'_{il;jk}&=&\Lambda'_{il;jk}+\frac{1}{2}\Lambda'_{il;i_{1}i_{2}}G_{i_{1}j_{1}}G_{i_{2}j_{2}}M'_{j_{1}j_{2};jk}.\label{BS3}
\end{eqnarray}
One may notice that there is another BS kernel $\Lambda'$, but we
should emphasize that $\Lambda$ and $\Lambda'$ belong to different
hierarchies and $\Lambda'$ is lower than $\Lambda$. However, the two
different BS kernels have the same symmetries, which will be
discussed more carefully in the following. As one can see, there are
three terms (separated by parentheses) on the right-hand side of
Eq.~(\ref{Lamb}). The second term corresponds to diagrams which are
two-particle reducible in the $t$ channel while 2PI in the other two
channels. The third term corresponds to those which are two-particle
reducible in the $u$ channel while 2PI in the other two channels.
The first term includes diagrams which are 2PI in all the three
channels. We can see that there are no diagrams which are
two-particle reducible in the $s$ channel, since this kind of
diagrams are prohibited by the renormalizability of the theory. As
for the diagrams included in the $\Lambda^{\mathrm{2PI}}$, which are
2PI in all the three channels, one can easily recognize that the
lowest order contribution to $\Lambda^{\mathrm{2PI}}$ is just the
bare vertex $\Lambda^{\mathrm{2PI}}_{0}$ as shown in Fig.~\ref{f4}.
We also show the next two order diagrams
$\Lambda^{\mathrm{2PI}}_{1}$ and $\Lambda^{\mathrm{2PI}}_{2}$ in
Fig.~\ref{f4}, which are four and five loops, respectively. In
Fig.~\ref{f4} we use ``Perms'' to indicate all other possible
permutations of external legs. One can find that there is no
subdivergence in $\Lambda^{\mathrm{2PI}}_{1}$ in four dimensions,
and we only need a counterterm (here denoted as $\delta\lambda_{1}$)
to absorb the overall divergences. Different from the four-loop
diagrams, the $\Lambda^{\mathrm{2PI}}_{2}$ has subdivergences which
must be absorbed. For example, the second diagram in the second line
of Fig.~\ref{f4} is employed to absorb the subdivergences, and we
also notice that these diagrams are also 2PI in all channels. After
the subdivergences are canceled, we only need a counterterm
$\delta\lambda_{2}$ to absorb the overall divergences once more.
Continuing this procedure, finally we can write the first term on
the right-hand side of Eq.~(\ref{Lamb}) as
\begin{equation}
\Lambda^{\mathrm{2PI}}-i\delta\lambda
=\Lambda^{\mathrm{2PI}}_{f}-i\Delta\lambda,\label{Lam2PI}
\end{equation}
where $\Lambda^{\mathrm{2PI}}_{f}$ is not only 2PI in all channels
but also finite. $\Delta\lambda$ is related with $\delta\lambda$
through the following equation:
\begin{equation}
\delta\lambda
=\Delta\lambda+\delta\lambda_{1}+\delta\lambda_{2}+\cdots.\label{}
\end{equation}
Here $\Delta\lambda$ is kept for the following use.

It is easily checked that the kernel $\Lambda$ in Eq.~(\ref{Lamb})
has the same symmetry as $\Lambda'$:
$\Lambda_{ij;kl}=\Lambda_{ij;lk}=\Lambda_{ji;kl}=\Lambda_{kl;ij}$.
It is more convenient to discuss the renormalization in momentum
space. In momentum space, we attach momenta $p$, $-p$, $-k$, $k$
with external legs whose subscripts are $i$, $j$, $k$, $l$,
respectively. Then, the BS equations in Eq.~(\ref{BS2}) can be
written as
\begin{eqnarray}
M'(p,-k;-p,k)&=&\Lambda'(p,-k;-p,k)+\frac{1}{2}\int\frac{d^{4}q}{(2\pi)^{4}}\Lambda'(p,-k;-q,q+k-p)\nonumber \\
&&\times G(q)G(q+k-p)M'(q,p-q-k;-p,k). \label{BSmomen1}
\end{eqnarray}
Here the kernel $\Lambda'$ can be obtained by the corresponding 2PI
vacuum diagrams, same as that of the conventional BS equation, which
also needs to be truncated. For example, we can use 2PI vacuum
diagrams up to two-loop or three-loop to obtain $\Lambda'$, which
corresponds to truncate $\Lambda'$ to bare vertex or one-loop
four-point function, respectively. Certainly, if we truncate the 2PI
vacuum diagrams to higher order, we can obtain the four-point
diagrams which are 2PI in all the three channels, just like the
diagrams in Fig.~\ref{f4}. In fact, the kernel $\Lambda'$ can also
have the same structure as the $\Lambda$ in Eq.~(\ref{Lamb}), which
receives contributions from three different kinds of diagrams. As we
have emphasized above, due to many people's efforts and
contributions, we have learned that the BS equation in
Eq.~(\ref{BSmomen1}) can be renormalized, only when the kernel
$\Lambda'$ is 2PI in the channel in which the resummation is
performed through the BS equation, which is guaranteed by the 2PI
property of the vacuum graphs~\cite{van2002,Blaizot2004,Berges2005}.
In order to make $M'$ being finite, we need a counterterm in the
kernel $\Lambda'$. Therefore, we can express $\Lambda'$ as a sum of
a finite part and a divergent constant, i.e.,
\begin{equation}
\Lambda'(p,-k;-p,k)=\Lambda'_{f}(p,-k;-p,k)-i\Delta\lambda_{t},\label{}
\end{equation}
where the subscript $t$ indicate the $t$ channel. In the same way,
the BS equation in Eq.~(\ref{BS3}) in momentum space is
\begin{eqnarray}
M'(p,k;-p,-k)&=&\Lambda'(p,k;-p,-k)+\frac{1}{2}\int\frac{d^{4}q}{(2\pi)^{4}}\Lambda'(p,k;-q,q-k-p)\nonumber \\
&&\times G(q)G(k+p-q)M'(q,k+p-q;-p,-k). \label{BSmomen2}
\end{eqnarray}
Also, we write the kernel as
\begin{equation}
\Lambda'(p,k;-p,-k)=\Lambda'_{f}(p,k;-p,-k)-i\Delta\lambda_{u},\label{}
\end{equation}

Employing
$\Delta\lambda=\Delta\lambda_{s}+\Delta\lambda_{t}+\Delta\lambda_{u}$,
we express Eq.~(\ref{Lamb}) in momentum space as
\begin{eqnarray}
\Lambda(p,-p;-k,k)&=&\Lambda^{\mathrm{2PI}}_{f}(p,-p;-k,k)-i\Delta\lambda_{s}\nonumber \\
&&+M'(p,-k;-p,k)-\Lambda'_{f}(p,-k;-p,k)\nonumber \\
&&+M'(p,k;-p,-k)-\Lambda'_{f}(p,k;-p,-k). \label{Eq24}
\end{eqnarray}
One can see that $\Lambda$ is also a sum of a finite part and a
divergent constant ($\Delta\lambda_{s}$), which make the four-point
vertex $M$ in the following BS equation finite:
\begin{eqnarray}
M(p,-p;-k,k)&=&\Lambda(p,-p;-k,k)+\frac{1}{2}\int\frac{d^{4}q}{(2\pi)^{4}}\Lambda(p,-p;-q,q)\nonumber \\
&&\times G^{2}(q)M(q,-q;-k,k). \label{BSmomen3}
\end{eqnarray}
We should emphasize that diagrams included in the kernel $\Lambda$
through resummation by Eq.~(\ref{BSmomen1}) and Eq.~(\ref{BSmomen2})
are 2PI in the $s$ channel, which guarantees that
Eq.~(\ref{BSmomen3}) can be renormalized.

Here we give some examples. The simplest case is that the kernel
$\Lambda'$ in Eq.~(\ref{BSmomen1}) is just the bare vertex. Then
$M'$ is only dependent on $p-k$ and Eq.~(\ref{BSmomen1}) can be
simplified to
\begin{eqnarray}
M'(p-k)&=&-i(\lambda+\Delta \lambda_{t})+\frac{1}{2}[-i(\lambda+\Delta \lambda_{t})]M'(p-k)\int\frac{d^{4}q}{(2\pi)^{4}}\nonumber \\
&&\times G(q)G(q+k-p). \label{}
\end{eqnarray}
Dividing both sides of the equation with
$-i(\lambda+\Delta\lambda_{t})M'(p-k)$ and using the free propagator
$G_{0}$, we obtain
\begin{eqnarray}
\frac{1}{-i(\lambda+\Delta\lambda_{t})}&=&\frac{1}{M'(p-k)}\nonumber \\
&&-\frac{i}{2(4\pi)^{2}}\Big(\frac{1}{\epsilon}+\int_{0}^{1}dx\ln\frac{\bar{\mu}^{2}}{m^{2}-x(1-x)(p-k)^{2}}\Big),\label{}
\end{eqnarray}
where we have employed the dimensional regularization and
$\bar{\mu}^{2}=4\pi\mu^{2}e^{-\gamma_{E}}$ is a mass scale. If we
introduce the following renormalization condition:
\begin{equation}
\frac{1}{-i(\lambda+\Delta\lambda_{t})}=\frac{1}{-i\lambda}-\frac{i}{2(4\pi)^{2}}\frac{1}{\epsilon},\label{}
\end{equation}
i.e.,
\begin{equation}
\Delta\lambda_{t}=\frac{\lambda}{1-\frac{\lambda}{2(4\pi)^{2}}\frac{1}{\epsilon}}-\lambda
=\lambda\sum_{n=1}^{\infty}\Big(\frac{\lambda}{2(4\pi)^{2}}\frac{1}{\epsilon}\Big)^{n}.\label{}
\end{equation}
Then we have
\begin{equation}
M'(p-k)=\frac{-i\lambda}{1+\frac{\lambda}{2(4\pi)^{2}}\int_{0}^{1}dx\ln\frac{\bar{\mu}^{2}}{m^{2}-x(1-x)(p-k)^{2}}}.\label{}
\end{equation}
Finally, we get the kernel $\Lambda$ as given by
\begin{eqnarray}
\Lambda(p,-p;-k,k)&=&i\lambda+\frac{-i\lambda}{1+\frac{\lambda}{2(4\pi)^{2}}\int_{0}^{1}dx\ln\frac{\bar{\mu}^{2}}{m^{2}-x(1-x)(p-k)^{2}}}
\nonumber \\
&&+\frac{-i\lambda}{1+\frac{\lambda}{2(4\pi)^{2}}\int_{0}^{1}dx\ln\frac{\bar{\mu}^{2}}{m^{2}-x(1-x)(p+k)^{2}}}-i\Delta\lambda_{s},\label{}
\end{eqnarray}
where $\Lambda^{\mathrm{2PI}}$ in Eq.~(\ref{Eq24}) is also truncated
to bare vertex. One can easily find that the kernel $\Lambda$ has
the asymptotic behavior:
$\Lambda(p,-p;-k,k)-\Lambda(\tilde{p},-\tilde{p};-k,k)\sim 1/k$ at
large $k$, which is the key point to renormalize the BS equation in
Eq.~(\ref{BSmomen3}).

\begin{figure}[!htb]
\includegraphics[scale=0.65]{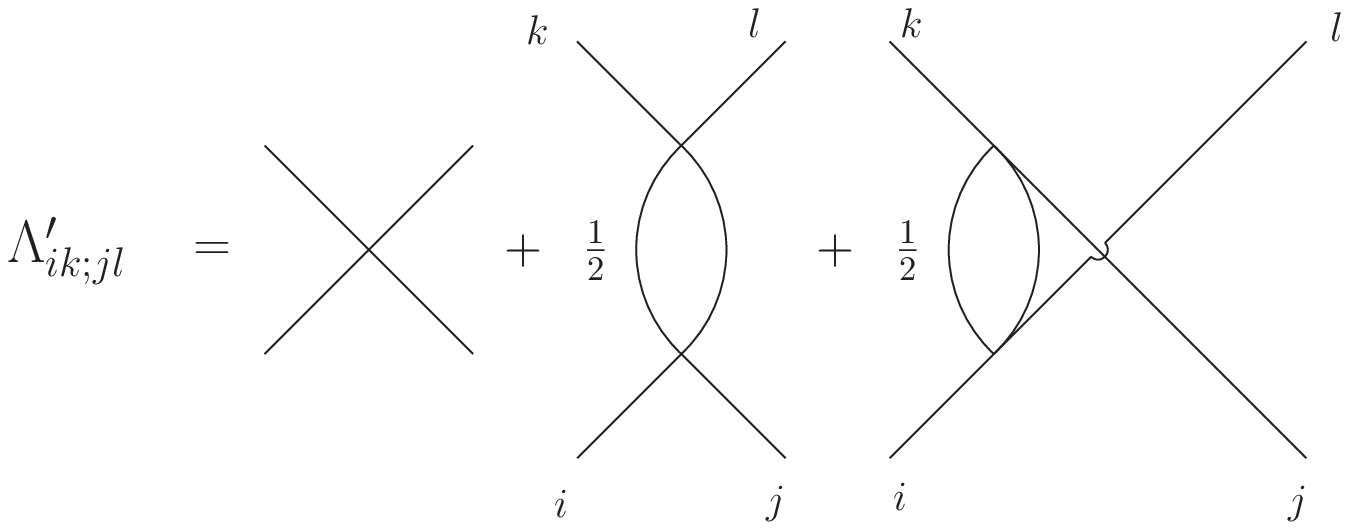}
\caption{Contributions to the kernel $\Lambda'_{ik;jl}$ up to
one-loop.}\label{f1}
\end{figure}

\begin{figure}[!htb]
\includegraphics[scale=0.5]{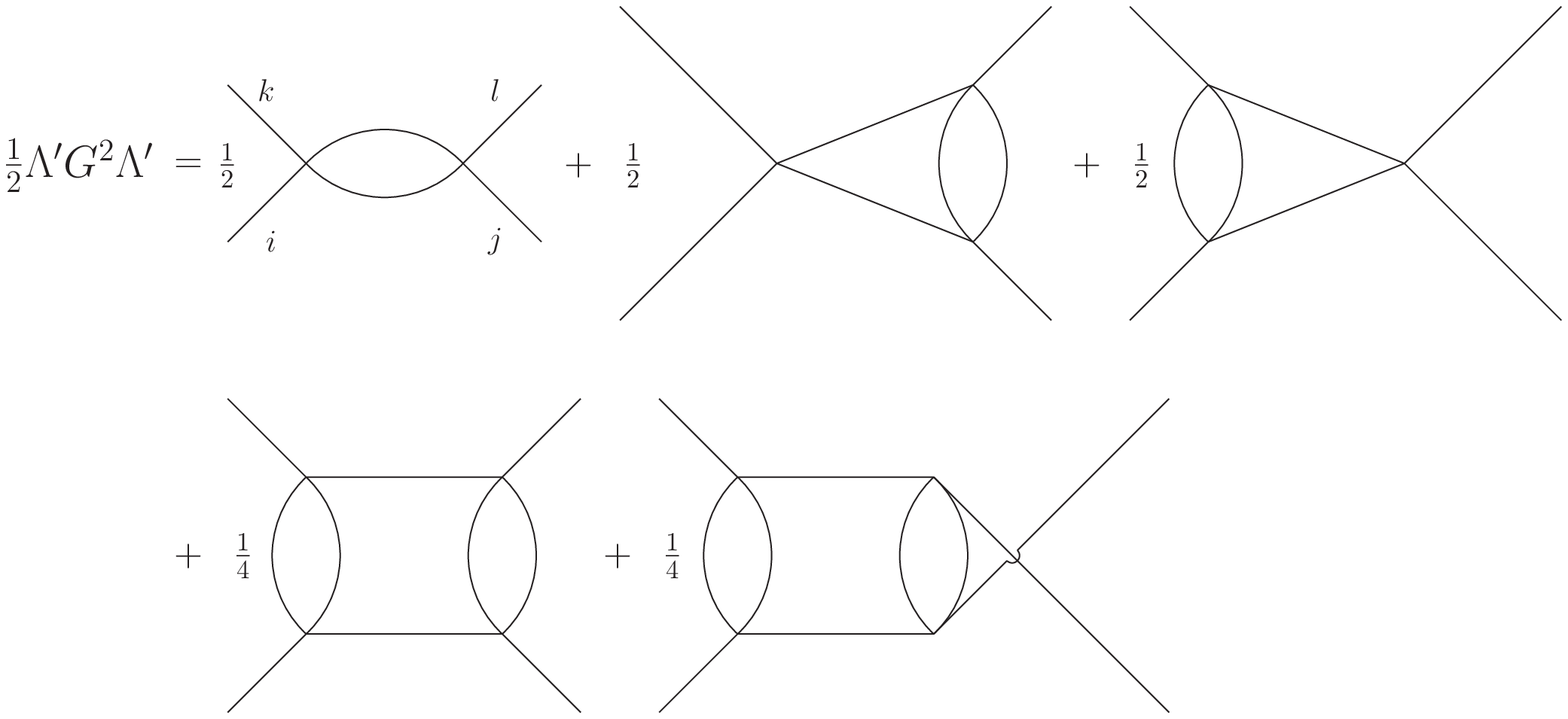}
\caption{Diagrams included in the second term on the right hand side
of Eq.~(\ref{Mprimexp}).}\label{f2}
\end{figure}

\begin{figure}[!htb]
\includegraphics[scale=0.5]{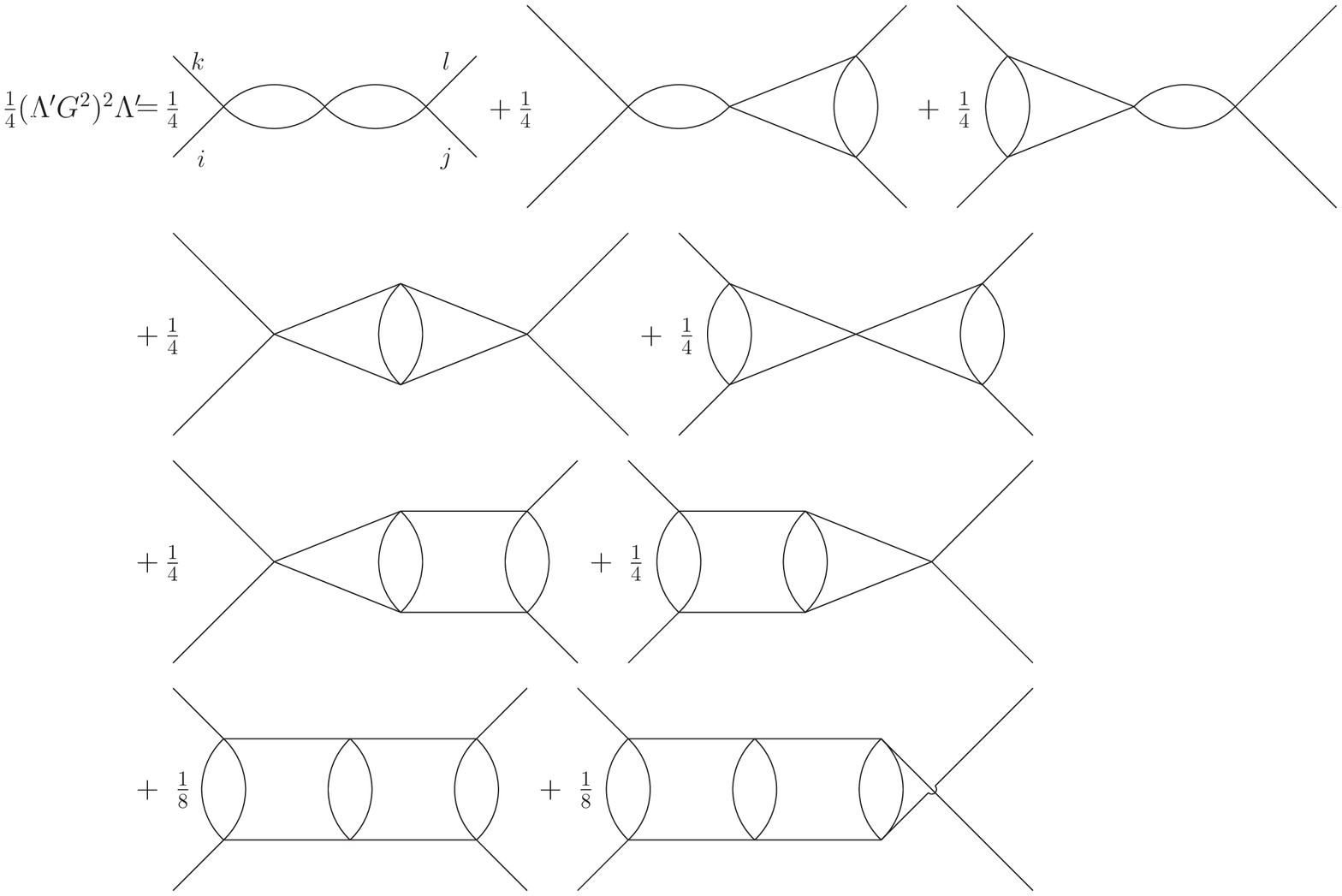}
\caption{Diagrams included in the third term on the right hand side
of Eq.~(\ref{Mprimexp}).}\label{f3}
\end{figure}

Then we consider a more complicated case and include one-loop
contributions to the kernel $\Lambda'$ in Eq.~(\ref{BS2}), which is
shown in Fig.~\ref{f1}. From the BS equation in Eq.~(\ref{BS2}), one
can rewrite $M'$ as
\begin{equation}
M'=\Lambda'+\frac{1}{2}\Lambda'G^{2}\Lambda'+\frac{1}{4}\Lambda'G^{2}\Lambda'G^{2}\Lambda'+....,\label{Mprimexp}
\end{equation}
where the subscripts are not labeled explicitly. One can see that
the second and third terms on the right hand side of the equation
above correspond to one and two iterations. Substituting the kernel
$\Lambda'$ in Fig.~\ref{f1}, we get their diagram representations
which are shown in Fig.~\ref{f2} and Fig.~\ref{f3}, respectively.
One can observe that all diagrams in Fig.~\ref{f2} and Fig.~\ref{f3}
are 2PI in the $s$ channel, which confirms the renormalizability of
the BS equation in Eq.~(\ref{BSmomen3}).


In fact, we could generalize our approach presented above. One can
use $M$ in Eq.~(\ref{BS1}) to construct the kernel of another BS
equation, for example,
\begin{equation}
\bar{\Lambda}_{ij;kl}=(\Lambda_{ij;kl}^{\mathrm{2PI}}-i\delta\bar{\lambda}\delta_{ij}\delta_{ik}\delta_{il})
+(M_{ik;jl}-\Lambda_{ik;jl})+(M_{il;jk}-\Lambda_{il;jk}),\label{Lambar}
\end{equation}
which has the same structure as Eq.~(\ref{Lamb}).  Then this new BS
equation with kernel $\bar{\Lambda}$ and the BS equations discussed
above constitute a set of BS equations with three different
hierarchies. One can easily check that these BS equations of
different hierarchies will not result in an over-counting of
diagrams. This is because Eq.~(\ref{Lamb}) and Eq.~(\ref{Lambar})
guarantee that the kernels are 2PI in the channel in which the
summations are performed through the BS equations. This is also the
reason why all the BS equations can be renormalized. This is the
hierarchy structure of the BS equations.

\section{Propagator and Effective Action}
\label{propagatorsection}

We have obtained the four-point vertices in the section above, and
the approximations are made on the level of the four-point
functions. Then the next task is to find the corresponding two-point
function, i.e., the propagator which constitutes the self-consistent
equations, together with the BS equations described above. In the
following, we will employ the functional renormalization group to
obtain the propagator.

Employing the effective action in Eq.~(\ref{Gamma_2PI}), taking
functional derivative  with respect to the propagator and
considering the stationary condition, we have
\begin{equation}
\frac{\delta \Gamma_{\kappa}[G]}{\delta
G}\bigg|_{G=G_{\kappa}}=0,\label{stati}
\end{equation}
which is in fact the gap equation as follows:
\begin{equation}
G_{\kappa}^{-1}=G_{0\kappa}^{-1}-\Sigma[G_{\kappa}],\label{Gap}
\end{equation}
where the self-energy $\Sigma$ is given by
\begin{equation}
\Sigma[G_{\kappa}]=2i\frac{\delta \Gamma_{\mathrm{int}}[G]}{\delta
G}\bigg|_{G=G_{\kappa}}.\label{}
\end{equation}
Taking derivative with respect to the flow parameter $\kappa$ on
both sides of the gap equation in Eq.~(\ref{Gap}), one arrives at
\begin{equation}
\partial_{\kappa}G_{\kappa}^{-1}=\partial_{\kappa}G_{0\kappa}^{-1}-\partial_{\kappa}\Sigma[G_{\kappa}].\label{pGap}
\end{equation}
The first term is
\begin{equation}
\partial_{\kappa}G_{0\kappa}^{-1}=i\partial_{\kappa}R_{\kappa}.\label{term1}
\end{equation}
The second term can be re-expressed as
\begin{eqnarray}
\frac{\partial \Sigma[G_{\kappa}]}{\partial \kappa}&=&\frac{\delta
\Sigma[G]}{\delta
G}\bigg|_{G=G_{\kappa}}\frac{\partial G_{\kappa}}{\partial \kappa}\nonumber\\
&=&-\frac{\delta \Sigma[G]}{\delta
G}\bigg|_{G=G_{\kappa}}G_{\kappa}\frac{\partial
G^{-1}_{\kappa}}{\partial \kappa}G_{\kappa}\nonumber\\
&=&-2i\frac{\delta^{2} \Gamma_{\mathrm{int}}[G]}{\delta G \delta
G}\bigg|_{G=G_{\kappa}}G_{\kappa}\frac{\partial
G^{-1}_{\kappa}}{\partial \kappa}G_{\kappa}\nonumber\\
&=&-\frac{1}{2}\Lambda[G_{\kappa}]G_{\kappa}\frac{\partial
G^{-1}_{\kappa}}{\partial \kappa}G_{\kappa}.\label{term2}
\end{eqnarray}
Substituting Eqs.~(\ref{term1})~(\ref{term2}) into Eq.~(\ref{pGap}),
one finds
\begin{equation}
\partial_{\kappa}G_{\kappa}^{-1}=i\partial_{\kappa}R_{\kappa}+\frac{1}{2}\Lambda[G_{\kappa}]G_{\kappa}^{2}\partial_{\kappa}
G^{-1}_{\kappa}.\label{pGap2}
\end{equation}
Employing the BS equation in Eq.~(\ref{BS1}), we can solve
$\partial_{\kappa}G_{\kappa}^{-1}$ in Eq.~(\ref{pGap2}) and arrive
at
\begin{equation}
\partial_{\kappa}G_{\kappa}^{-1}=i\partial_{\kappa}R_{\kappa}+\frac{i}{2}M[G_{\kappa}]G^{2}_{\kappa}\partial_{\kappa}
R_{\kappa},\label{propagator}
\end{equation}
which is the flow equation for the propagator. Blaizot \textit{et
al.} derived this equation firstly and found that this flow equation
is completely equivalent with the gap equation~\cite{Blaizot2011}.
We should emphasize that in the conventional approaches where the
approximations are made in the level of the effective potential, the
gap equation is easily obtained, so the flow equation for the
propagator seems to be not important. However, in our case the
kernel of the BS equation is resummed to infinite order through
another BS equation, and it is difficult to obtain the gap equation
directly. Therefore, we have to resort to the flow equation in
Eq.~(\ref{propagator}). Then the BS
equations~(\ref{BSmomen1}),~(\ref{BSmomen2}),~(\ref{BSmomen3}), and
the differential equation~(\ref{propagator}) constitute a closed
system, which can be solved.

Since the propagator is obtained, the effective action can also be
found. Differentiating the effective action in Eq.~(\ref{Gamma_2PI})
with respect to the flow parameter $\kappa$, one finds
\begin{eqnarray}
\partial_{\kappa}\Gamma_{\kappa}[G_{\kappa}]&=&
\frac{\partial\Gamma_{\kappa}[G]}{\partial\kappa}\bigg|_{G=G_{\kappa}}
+\frac{\partial\Gamma_{\kappa}[G]}{\partial
G}\bigg|_{G=G_{\kappa}}\frac{\partial
G_{\kappa}}{\partial\kappa}\nonumber\\
&=&\frac{\partial\Gamma_{\kappa}[G]}{\partial\kappa}\bigg|_{G=G_{\kappa}}\nonumber\\
&=&-\frac{1}{2}\mathrm{Tr}[(\partial_{\kappa}R_{\kappa})G_{\kappa}],\label{}
\end{eqnarray}
where we have used Eq.~(\ref{stati}) in the second line. This is the
flow equation for the effective action first derived in
Ref.~\cite{Wetterich1993}. Integrating the flow equation, one can
get the effective action.

\section{Summary and Outlook}
\label{summary}

In this work, we have proposed a new resummation scheme under the
formalism of 2PI effective action theory. We employ the hierarchy
structure of the BS equations to greatly improve on the
approximations for the BS kernel. Resumming the kernel in $t$ and
$u$ channels to infinite order is equivalent to truncate the
effective action to infinite order. Furthermore, our approximation
approaches do not violate the renormalizability of the theory, which
is very important for numerical calculations. We also obtain the
two-point function from the four-point one through flow evolution
equations. Therefore, BS equations of different hierarchies and the
flow evolution equation for the propagator constitute a closed
system, which can be solved completely.

We should note that there are two hierarchies of BS equations in our
approach. Therefore, comparing the conventional BS equations, we
need more computer time in our calculations. But because of
remarkable increases in computer power made possible by clusters,
this problem is not difficult to solved.

Since we have proposed to employ the hierarchy structure of the BS
equations to improve on the approximations for the BS kernel, it is
very interesting to apply our approaches to detailed problems. One
potential interesting problem is to compute the shear viscosity of
the quark--gluon plasma or other thermal fields. As we know the
shear viscosity is very sensitive to the properties of the
four-point vertex, so it is expected that our approach will advance
the computation of the shear viscosity. Furthermore, our approaches
can also be applied to strongly correlated electron systems, for
example, our approaches can be used to study the two-dimensional
Hubbard model, which may shed new light on our understanding about
high $T_{c}$ superconductors.

\begin{acknowledgements}
I am indebted to M.E. Carrington for useful discussions. This work
was supported by the National Natural Science Foundation of China
under Contracts No. 11005138.
\end{acknowledgements}



\end{document}